\documentclass[journal=jacsat,manuscript=article]{achemso}

\usepackage[version=3]{mhchem} 
\usepackage{textgreek}
\usepackage{gensymb}
\usepackage{amsmath}

\author{Christopher G. Bailey}
\affiliation[University of Southampton]
{Physics and Astronomy department, University of Southampton, University Road, Southampton, SO17 1BJ}
\email{C.Bailey@Soton.ac.uk}
\author{Giacomo M. Piana}
\affiliation[University of Southampton]
{Physics and Astronomy department, University of Southampton, University Road, Southampton, SO17 1BJ}
\author{Pavlos G. Lagoudakis}
\affiliation[University of Southampton]
{Physics and Astronomy department, University of Southampton, University Road, Southampton, SO17 1BJ}
\alsoaffiliation[Skoltech]
{Skolkovo Institute of Science and Technology (Skoltech), Skolkovo, Moscow, Russia, 121205}

\title[An \textsf{achemso} demo]
  {High-energy optical transitions and optical constants of \ce{CH3NH3PbI3} measured by spectroscopic ellipsometry and spectrophotometry}

\abbreviations{IR,NMR,UV}

\begin{document}



\begin{abstract}
  Optoelectronics based on metal halide perovskites (MHPs) have shown substantial promise, following more than a decade of research. For prime routes of commercialization such as tandem solar cells, optical modeling is essential for engineering device architectures, which requires accurate optical data for the materials utilized. Additionally, a comprehensive understanding of the fundamental material properties is vital for simulating the operation of devices for design purposes. In this article, we use variable angle spectroscopic ellipsometry (SE) to determine the optical constants of \ce{CH3NH3PbI3} (MAPbI$_3$) thin films over a photon energy range of 0.73 to 6.45 eV. We successfully model the ellipsometric data using six Tauc-Lorentz oscillators for three different incident angles. Following this, we use critical-point analysis of the complex dielectric constant to identify the well-known transitions at 1.58, 2.49, 3.36 eV, but also additional transitions at 4.63 and 5.88 eV, which are observed in both SE and spectrophotometry measurements. This work provides important information relating to optical transitions and band structure of MAPbI$_3$, which can assist in the development of potential applications of the material.
\end{abstract}

\section{Introduction}

Metal halide perovskites (MHPs) are a propitious group of materials for thin-film optoelectronics, with the research field still thriving and record solar cell efficiencies now reaching greater than 25\%.\cite{NREL2019} As photovoltaic devices begin to approach the theoretical efficiency limit for a single junction \cite{Correa-Baena2017,Sha2015,Polman2016,Braly2018}, fine-tuning of the device structure is required for further optimization. This can only be feasible with the assistance of optical simulations, which require accurate knowledge of the optical properties of device layers. Moreover, one of the most promising applications for commercialization of MHPs are tandem solar cells, where the design also requires extensive optical modeling. \cite{Paetzold2017,Albrecht2016,Jiang2016,Mantilla-Perez2017} The archetypal material for studying the properties of MHPs is \ce{CH3NH3PbI3} (MAPbI$_3$), which has been characterised scrupulously, revealing remarkable fundamental features for optoelectronics including high absorption coefficients ($\sim 10^4$--$10^5$ cm$^{-1}$ at 2 eV)\cite{Shirayama2016,DeWolf2014,Xing2013,Sun2014}, low exciton binding energies \cite{Miyata2015,Even2014,yang2015,Piana2019}, and small charge-carrier effective masses. \cite{Miyata2015,Giorgi2013,Ziffer2016} As a result, many other optoelectronic applications of MHPs have been developed, including lasers \cite{Li2018a,Wang2018,Wang2016,Stylianakis2019,Liao2015,Zhu2015}, light emitting diodes (LEDs) \cite{Tan2014,Cho2015,Ling2016,Liang2016} and photodetectors \cite{Saidaminov2015,Fang2015a,Bao2017}.

Many studies have also been made to characterize the optical constants of MAPbI$_3$ thin films \cite{Ball2015,Ziang2015,Jiang2015,Marronnier2018,Wang2019,Lin2015,Leguy2015,Leguy2016a,Loper2015,Shirayama2016,Jiang2016a,VanEerden2017,Demchenko2016,Ndione2016,Guerra2017}, and also other MHPs \cite{Brittman2016,Chen2019,Shirai2017,Zhao2018,Alias2016,Werner2018,Whitcher2018} with varying results and analysis protocols. Optical constants are typically determined by fitting data from spectroscopic ellipsometry (SE) measurements to a dispersion model, which can then be compared with results from spectrophotometry (reflectance and transmission spectroscopy). With spectrophotometry measurements on MHPs, data for wavelengths of $\lambda < 300$ nm (photon energies of $E > 4$ eV) is usually omitted. This may be due to a) limitations of the measurement range of the equipment used, b) strong absorption of glass substrates and/or c) strong absorption of the material (for thicker films and single crystals), with the latter two resulting in a weak (noisy) signal for optical transmission. Hence, optical transitions at these photon energies are seldom discussed. Nevertheless, studies of this part of the spectrum are important for validating theoretical work regarding the band structure and ultimately understanding the fundamental properties of MHPs. Often the focus of band-structure calculations is to reproduce the bandgap energy of MHPs, however an accurate model should predict spectral features at all photon energies observable by experiment.

Here we use variable angle SE to determine the optical constants of a MAPbI$_3$ thin film over a wide spectral range. The ellipsometric data is modeled using Tauc-Lorentz oscillators for measurements from three different incident angles. We report on several optical transitions which are observed at ultraviolet photon energies in both SE data and spectrophotometry measurements. We present a simple step-by-step approach that can be used for utilizing SE as a method for accurately determining the optical constants, thickness and surface roughness of MHP thin films.

\section{Results and discussion}

\begin{figure}[h]
\centering
  \includegraphics[height=9.5cm]{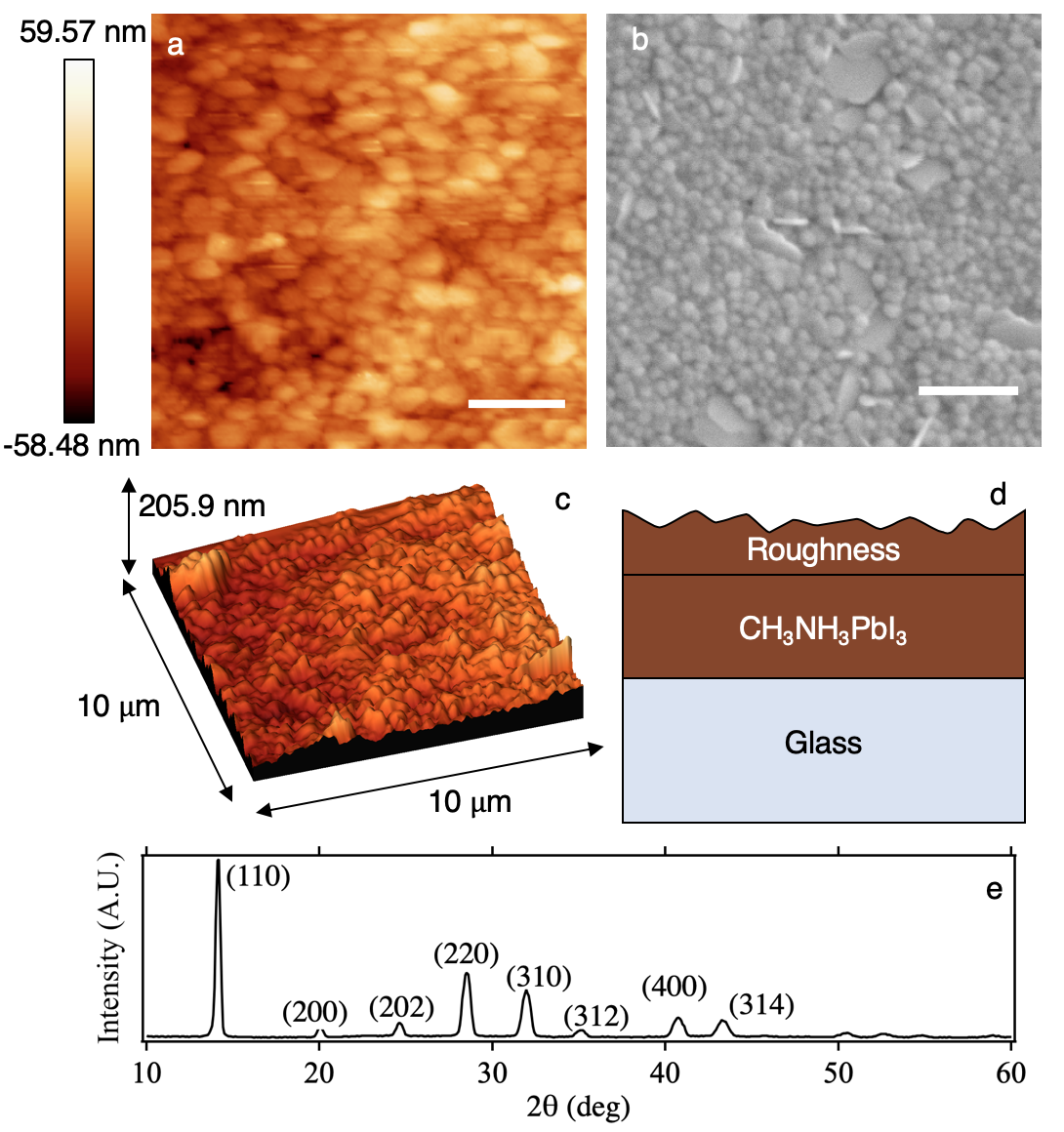}
  \caption{Characterization of the MAPbI$_3$ thin films: a) AFM image. b) Top-view SEM image. c) 3D AFM topography. d) Schematic representation of the optical model. e) X-ray diffraction pattern. Scale bars are 1 \textmu m.}
  \label{fgr:sample}
\end{figure}

We prepare MAPbI$_3$ via a one-step, gas-assisted spin coating method to achieve smooth uniform thin films.\cite{Conings2016} This technique utilizes the high pressure flow of inert gas on the surface of the film during the coating process, bypassing the need for anti-solvent quenching.\cite{Jeon2014a,Xia2016,Gao2018} We use a precursor of PbI$_2$ and methylammonium iodide (MAI) (1:1) in a mixed solvent solution of DMF and DMSO at a concentration of 1M. Figure \ref{fgr:sample} shows the sample morphology from atomic force microscopy (AFM) and scanning electron microscopy (SEM). The film has a relatively uniform surface with grain size of $\sim$ 100 to 200 nm.
We performed SE measurements on thin films of MAPbI$_3$ to obtain the $\Psi$ and $\Delta$ values over the spectral range of 192 to 1696 nm, for three angles of incidence (65\degree, 70\degree, and 75\degree). The parameter $\Delta$ is the induced phase difference between the $s$- and $p$- polarized light and $\Psi$ is the angle whose tangent is given by the ratio of the magnitudes of the reflection coefficients, $|R_s|$ and $|R_p|$ for the $s$ and $p$ polarization planes, respectively, i.e. $\tan{\Psi}=|R_p|/|R_s|$. We ensure that the sample is measured immediately following exposure to ambient conditions, after being packaged in an inert environment so that minimal degradation occurs prior to measurement. Figure \ref{fgr:sample}e shows the X-ray diffraction pattern of the film in this condition, displaying a high degree of crystallinity and absence of peaks that indicate degradation. To model the ellipsometry results, we firstly use a Cauchy film model to fit the data in the transparent region ($>$ 800 nm), where oscillations due to interference are observed. The refractive index from the Cauchy dispersion model is given by:
\begin{equation}
  n(\lambda)=A+\frac{B}{\lambda^{2}}+\frac{C}{\lambda^{4}},
  \label{Eq:Cauchy}
\end{equation} 
where $A$, $B$ and $C$ are the Cauchy coefficients to be fitted, and $\lambda$ is the wavelength. This gives a good estimate for the thickness of the film, which is not fitted during the next phase. Subsequently, the Cauchy film model is parameterized to convert it to a basis-spline (b-spline) model which is used to fit the transparent region of the spectrum. The fit is then expanded to wavelengths below the bandgap, where the film is absorbing. The b-Spline model is then forced to be consistent with Kramers-Kronigs relations and fit to the data once again with updated parameters. Finally, the b-Spline model is parameterized to build an oscillator model with six Tauc-Lorentz oscillators. In the Tauc-Lorentz model, the imaginary part of the dielectric constant is given by 
\begin{equation}
\varepsilon_{2}=\left\{\begin{array}{ll}{\sum_{i=1}^{N} \frac{1}{E} \cdot \frac{A_{i}  E_{i}  B_{i} \left(E-E_{\mathrm{G}}\right)^{2}}{\left(E^{2}-E_{i}^{2}\right)^{2}+B_{i}^{2}  E^{2}}} & {\text { for } {E}>{E}_{\mathrm{G}}} \\ {0} & {\text { for } {E} \leq {E}_{\mathrm{G}}}\end{array}\right.
 \label{Eq:TL}
\end{equation}
Here, $E_\mathrm{G}$ is the bandgap, used as a coupled fitting parameter across all oscillators. The parameter $A_i$ is the amplitude and $C_i$ is the broadening of each oscillator peak centered at the energy $E_i$. Once the parameters for $\varepsilon_{2}$ are obtained, $\varepsilon_{1}$ is found using the analytic solution of the Kramers-Kronig integration. \cite{Jellison1996} The resulting model for pseudo-optical constants can be fitted to the data for $\Psi$ and $\Delta$ using
\begin{equation}
    \langle\tilde{\varepsilon}\rangle=\sin ^{2} \theta_{i}\left[1+\tan ^{2} \theta_{i}\left(\frac{1-\rho}{1+\rho}\right)^{2}\right],
    \label{Pseudo}
\end{equation}
where $\theta_{i}$ is the angle of incidence and $\rho$ is the induced polarization change ($\rho=\tan (\Psi) e^{i \Delta}$). The parameters are fitted globally for three angles of incidence (65\degree, 70\degree, and 75\degree) over the wavelength range of 192 to 1696 nm. We incorporated the film thickness and surface roughness in the model (Figure \ref{fgr:sample}d) and all parameters are updated to obtain the lowest possible MSE (mean squared error).

\begin{figure}[h]
\centering
  \includegraphics[height=8cm]{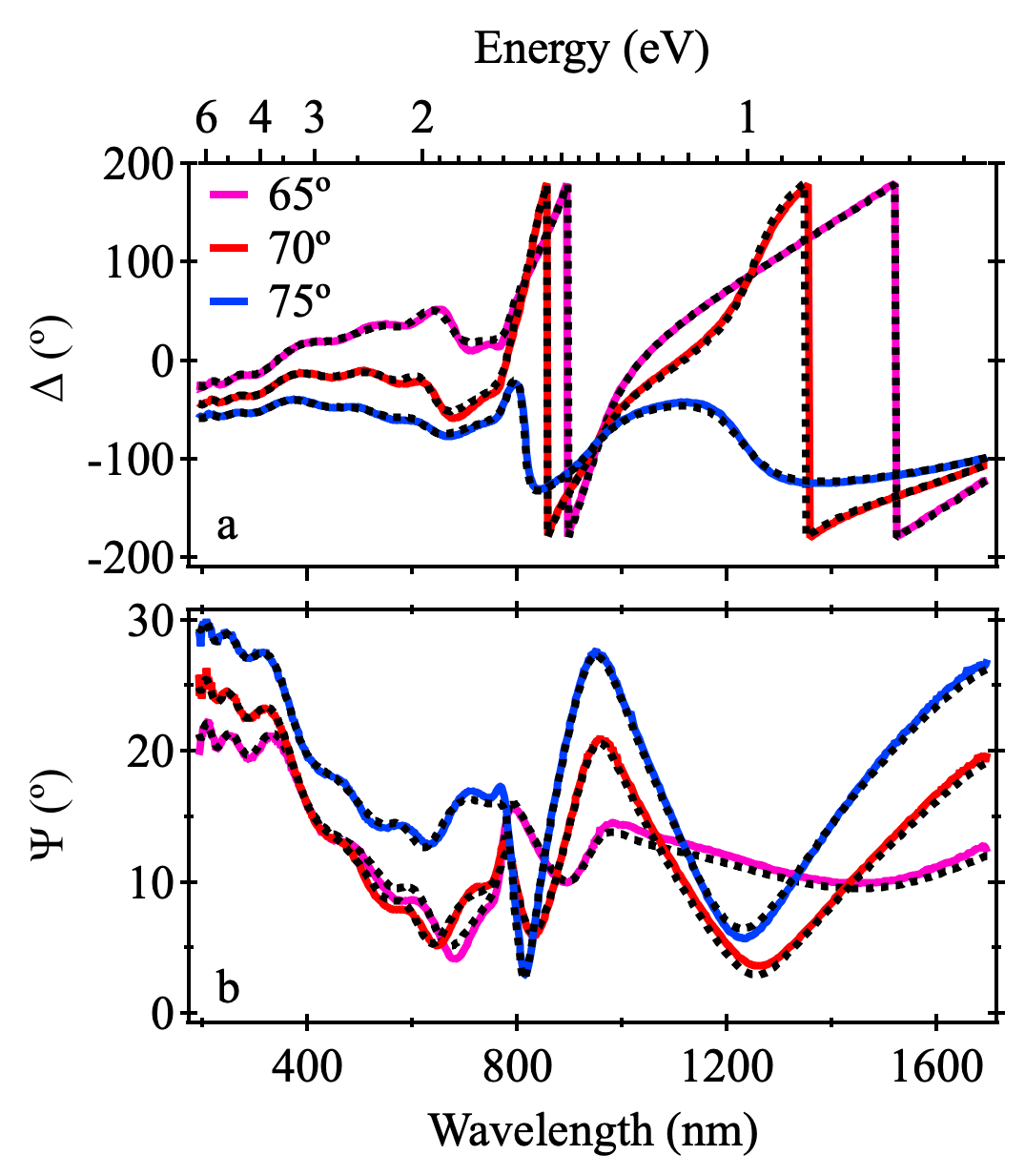}
  \caption{Polarization data obtained from variable angle SE measurements. The dashed lines show the fits from the Tauc-Lorentz model.}
  \label{fgr:psidel}
\end{figure}

The $\Psi$ and $\Delta$ data is shown in Figure \ref{fgr:psidel}, along with the fit from the Tauc-Lorentz oscillator model.
A MSE value of 12.50 is achieved, indicating excellent agreement between the model and data. We obtain a thickness value of 446.84 $\pm$ 0.24 nm, which is in agreement of the thickness obtained using profilometry (452 $\pm$20 nm). The roughness of the film is found to be $\sim$ 10 nm, which is also close to the RMS (root mean square) roughness value of 13 nm obtained by AFM (Figure \ref{fgr:sample}a and c). It is important to factor the surface roughness, since it has a substantial impact on the determination of optical characteristics of the film, with larger values causing erroneous interpretation \cite{Fujiwara2018}. The fitting parameters obtained from the Tauc-Lorentz model are outlined in Table 1.
\begin{table}[h]
\small
  \caption{\ Fitting parameters of the Tauc-Lorentz oscillators}
  \label{tbl:table1}
  \begin{tabular}{llll}
    \hline
    Oscillator & Center (eV) & $B_i$ (eV) & $A_i$ \\
    \hline
    $E_{0}$ ($E_{G}$) & 1.565 $\pm$ 0.0046 & 0.127 $\pm$ 0.0043 & 99.391 $\pm$ 5.8436\\
    $E_{1}$ & 2.535 $\pm$ 0.0173 & 0.918 $\pm$ 0.0637 & 13.381 $\pm$ 3.4237 \\
    $E_{2}$ & 3.390 $\pm$ 0.0057 & 0.948 $\pm$ 0.0334 & 4.931 $\pm$ 0.8600 \\
    $E_{3}$ & 4.665 $\pm$ 0.0183 & 1.300$\pm$ 0.0889 & 5.974 $\pm$ 0.6147  \\
    $E_{4}$ & 5.790 $\pm$ 0.0332 & 1.083 $\pm$ 0.1359 & 3.899 $\pm$ 0.8210 \\
    $E_{5}$ & 6.521 $\pm$ 0.0773 & 0.615 $\pm$ 0.2366 & 1.552 $\pm$ 0.8695 \\
    $E_\text{UV}$ & 7.674 $\pm$ 0.2404 &  & 46.300 $\pm$ 7.4674  \\
    \hline
  \end{tabular}
\end{table}

We find that six oscillators are required to minimize the MSE of the fit and reproduce the data accurately. If fewer than six oscillators are used (four or five, for example), the fit does not capture all of the features of the $\Delta$ and $\Psi$ curves (Figure \ref{fgr:psidel}a and b, respectively) and instead converges to oscillator parameters that provide an average value to account for several transitions. Only very slight reductions in the MSE can be obtained if more oscillators are used, and the amplitude of the additional oscillators have to be very low to achieve this. Therefore, we can conclude that any more or fewer than six main oscillators appears to create a non-physical model. A value of $\epsilon_{\infty} = 1.016 \pm 0.1$ is fitted to provide a dielectric constant at high energies, in order to prevent $\varepsilon_{1}$ from becoming zero following the application of the Kramers-Kronig transformation. The element $E_\text{UV}$ is also included as an unbroadened oscillator to account for high-energy absorption, which effectively provides a 'tilt' to the dielectric constant. The bandgap value of $E_{\mathrm{G}} = 1.565$ eV is determined as a coupled fitting parameter for all oscillators.

\begin{figure}[h]
\centering
  \includegraphics[height=10cm]{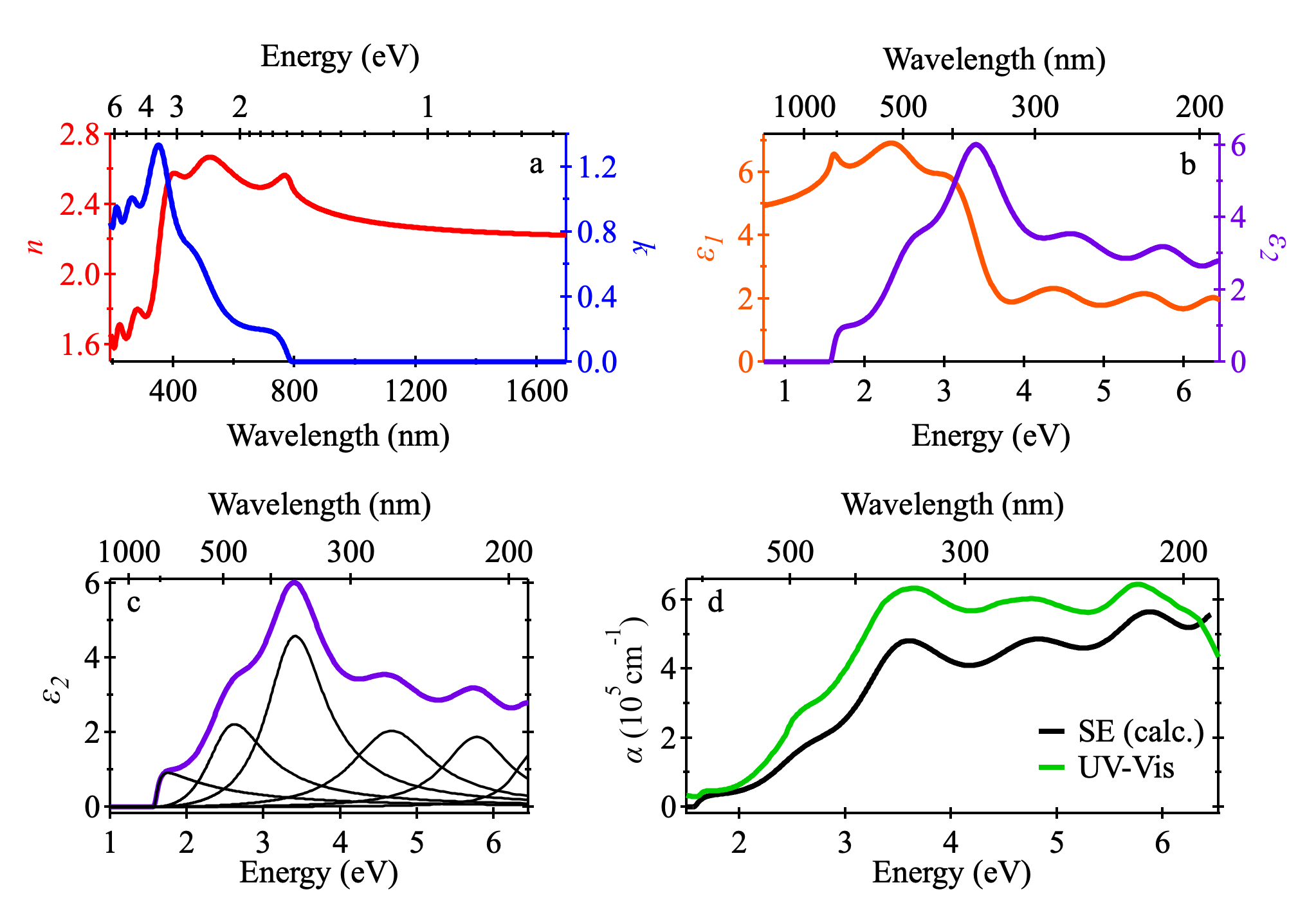}
  \caption{Optical constants determined from the SE data analysis: a) Refractive index ($n$) and extinction coefficient ($k$). b) Real ($\epsilon_1$) and imaginary ($\epsilon_2$) components of the dielectric constant. c) Tauc-Lorentz oscillators (black curves) used to model $\epsilon_2$ (purple curve). d) Absorption coefficient calculated from SE data (black) and estimated from spectrophotometry data (green). }
  \label{fgr:nk}
\end{figure}

The optical constants determined from the analysis are presented in Figure \ref{fgr:nk}.
The black curves in Figure \ref{fgr:nk}c show the individual contribution from the Tauc-Lorentz oscillators to $\varepsilon_{2}$. Figure \ref{fgr:nk}d shows the absorption coefficient, $\alpha$ derived from the extinction coefficient, $k$, calculated using the relation $\alpha=4 \pi k / \lambda$. We performed spectrophotometry measurements on a different MAPbI$_3$ thin film deposited on quartz to obtain an estimate for the absorption coefficient, which is overlaid with the calculated absorption coefficient in Figure \ref{fgr:nk}d. For these measurements, a thinner film was deposited using a 0.5M solution to obtain a thickness of $\sim$ 110 nm, to ensure the transmission signal was strong enough to be detected at highly absorbing wavelengths. 

The observed transitions in the absorbing region of the experimental $\alpha$ spectrum appear to corroborate with the modeled data from the ellipsometry measurements. For the SE-calculated absorption coefficient, the value at 2 eV is $\sim$ 4.6 x $10^4$ cm$^{-1}$, which is in good agreement with previous work.\cite{Loper2015,Jiang2015,Xing2013} The absolute amplitude of the absorption coefficient obtained from spectrophotometry should not be considered as an accurate value for comparison, since scattered light from the sample is not accounted for. Instead, we use the data to compare the shape and position of the optical transitions. One major difference between the two absorption coefficients is position and relative amplitude of the highest energy peak. In the spectrophotometry data, it appears that the two highest energy peaks coincide, creating a high energy shoulder at $\sim$ 6.3 eV in the peak observed with a maximum at $\sim$ 5.8 eV. In the SE data however, it appears that the highest energy peak is blueshifted in comparison, with the maximum lying outside of the measurement range, with the Tauc Lorentz oscillator associated with this peak centered at 6.52 eV (Table 1). 

\begin{figure}[h]
\centering
  \includegraphics[height=6cm]{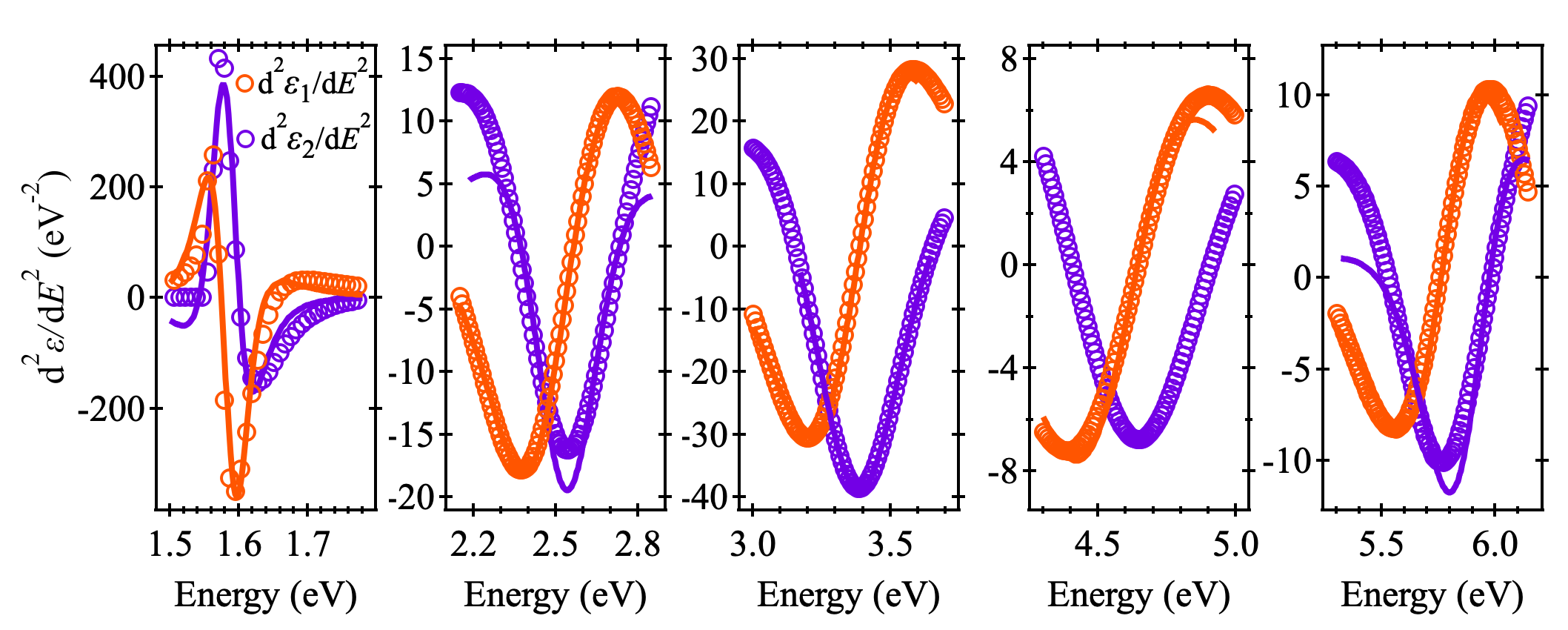}
  \caption{Critical-point analysis of the dielectric constant. The solid lines show the fitting of equation \ref{eq:CPFit}.}
  \label{fgr:cpfitting_V3}
\end{figure}

To investigate the optical transitions in MAPbI$_3$, we used critical-point (CP) analysis to accurately determine the transition energies. The CP model gives the following expression for the second derivative of the complex dielectric constant for any given excitonic transition, $j$:
\begin{equation}
\frac{\mathrm{d}^{2}\varepsilon}{\mathrm{d} E^{2}} = 2 A_{j} \mathrm{e}^{i \varphi_{j}}\left(E-E_{\mathrm{C}j}+i \Gamma_{j}\right)^{-3}.
\label{eq:CPFit}
\end{equation}
The energy of the transition is given by $E_{\mathrm{C}j}$, the broadening is $\Gamma_{j}$, the amplitude is $A_{j}$, the phase is $\varphi_{j}$, and $i$ is the imaginary unit. This equation is used to fit the real and imaginary parts of the dielectric function, as shown in Figure \ref{fgr:cpfitting_V3}.
The transition energies obtained from this analysis are $E_{\mathrm{C}0} (E_{\mathrm{G}})= 1.584 \pm 0.001$ eV, $E_{\mathrm{C}1} = 2.49 \pm 0.02$ eV, $E_{\mathrm{C}2} = 3.36 \pm 0.01$ eV, $E_{\mathrm{C}3} = 4.63 \pm 0.01$ eV $E_{\mathrm{C}4} = 5.88 \pm 0.01$ eV. We could not extract an energy associated with the transition at the high energy limit of the dispersion, since it is not fully captured by the measurement range. 

The three transitions $E_{\mathrm{C}0}$, $E_{\mathrm{C}1}$ and $E_{\mathrm{C}2}$ are commonly identified \cite{Loper2015,Jiang2016,Wang2019}, however, $E_{\mathrm{C}3}$ and $E_{\mathrm{C}4}$ are not often observed or remarked on. Our obtained value for the bandgap is similar to those previously determined \cite{Guerra2017,Loper2015,DeWolf2014,Shirayama2016}, which has been attributed to a direct transition at the R symmetry point of the Brillouin zone \cite{Leguy2016a,Shirayama2016}. Shirayama \textit{et al.} reported the analysis of SE measurements over a wide spectral range, where several weak transitions are observed at high energies \cite{Shirayama2016}. In their work, critical points are identified at 2.53 and 3.24 eV in the dielectric function for MAPbI$_3$ and are assigned to the direct optical transitions at the M and X points in the pseudocubic Brillouin zone, respectively. Leguy \textit{et al}. also performed SE measurements up to photon energies up to 5.5 eV and fitted the data to determine critical-point energies of 1.62, 2.55, 3.31, 4.55, and 10 eV for single crystals of MAPbI$_3$ but slightly different values of 1.61, 2.5, 2.85, 3.38, and 6.97 eV for thin films.\cite{Leguy2016a} In the study, the authors use quasi-particle self-consistent GW (QS-GW) calculations to derive optical constants up to photon energies of 10 eV, which are in good agreement with experimental data within in the spectral range of the measurements. In the calculated extinction coefficient spectrum, peaks are also predicted at $\sim$ 5.5 and 6.7 eV, which are not captured in their ellipsometry experiments, but are relatively close to the transitions observed in our experiments here.\cite{Leguy2016a} Hirasawa and co-workers also observed the transition at 4.8 eV in the absorption spectrum at a temperature of 4.2 K \cite{Hirasawa1994}. Ndione and colleagues observed the transitions at $\sim$ 4.6 and 5.8 eV in the dielectric constant from SE measurements, but only included the transition at 4.6 eV in the critical-point analysis.\cite{Ndione2016} Demchenko \textit{et al}. performed SE measurements on MAPbI$_3$ and extracted a dielectric constant that closely matches our obtained function, including the high energy features, likely originating from $\Gamma$ point transitions \cite{Demchenko2016}. Analysis of their band structure calculations showed that time-dependent hybrid functional calculations based on exchange-tuned HSE (TD-HSE), with spin-orbit coupling included in the calculation best reproduced the experimental data. Guerra and colleagues also observed the transitions at $\sim$ 4.6 and 5.8 eV via SE but did not measure spectrophotometry data beyond 5 eV. \cite{Guerra2017} In all the above studies, the highest energy optical transition is not investigated via spectrophotometry measurements in addition to SE. 

\subsection{Conclusions}

In conclusion, we have used spectroscopic ellipsometry (SE) and spectrophotometry to identify optical transitions in MAPbI$_3$ thin films and have determined the optical constants over a wide spectral range. We have shown how the SE data can be fitted with a Tauc-Lorentz dispersion model with six oscillators, using parameters that can be physically justified. These analyses are critical for constructing optical models for use in designing structures such as tandem solar cells. We have used critical-point analysis to show transitions at 1.58, 2.49, 3.36, 4.63 and 5.88 eV from the SE data, which is in excellent agreement with results from spectrophotometry. The transitions at 4.63 and 5.88 eV are not usually investigated due to experimental limitations and provide important information relating to the band structure of MAPbI$_3$ and the potential applications of the material. Their origin remains unclear and further theoretical work should be carried out to understand the nature of these transitions and assist in accurately determining the fundamental physical characteristics of metal halide perovskites.

\section{Experimental}
\subsection{Sample Preparation}

Glass substrates were cleaned by sonicating in Helmanex III solution with dionized water, and subsequently sonicating in isopropyl alcohol. They are then treated with oxygen plasma for 10 minutes before being transferred to an inert environment for deposition. For the MAPbI$_3$ precursor solution, a mixed solvent system was used with DMF and dimethylsulfoxide (DMSO). PbI$_2$ and MAI were combined with a molar ratio of 1:1 in a DMF:DMSO ($\sim$ 9:1) solvent at a concentration of 1M to form a PbI$_2$-DMSO complex. The MAPbI$_3$ solution was dispensed on the glass substrate and spin coated at 3000 RPM and $\sim 10 s$ after the the procedure is initiated, a flow of nitrogen gas is applied to the film to dry the parent solvent and produce smooth films. The MAPbI$_3$ films are then annealed for 10 mins at 100 \degree C in a nitrogen atmosphere. For spectrophotometry measurements, a quartz substrate is used instead of glass and the precursor concentration of 0.5M is used instead to achieve a film thickness of $\sim$ 110 nm.

\subsection{Sample Characterization}
The samples were characterized using a Rigaku SmartLab X-ray diffractometer with a Cu K-alpha 9 kW anode. The scanning electron microscopy (SEM) was carried out using a Zeiss (LEO) 1450VP Scanning Electron Microscope with an accelerating voltage of 20 kV. The atomic-force microscopy (AFM) was performed with a Nanonics Imaging Hydra BioAFM system.

\subsection{Spectroscopic Ellipsometry}

A J.A. Woolam M-2000 spectroscopic ellipsometer was used to measure the induced polarization change in the MAPbI$_3$ thin films for incident angles of 65 \degree, 70 \degree and 75 \degree. The samples were packaged in an inert environment and measured immediately after opening to ambient conditions. CompleteEase (J.A. Woolam) software is then used for the data analysis and dispersion model fitting, as described in the main text. The root mean squared error is calculated using the equation: 

\begin{equation}
M S E=\sqrt{\frac{1}{3 n-m} \sum_{i=1}^{n}\left[\left(N_{Exp_{i}}-N_{Mod_{i}}\right)^{2}+\left(C_{Exp_{i}}-C_{Mod_{i}}\right)^{2}+\left(S_{Exp_{i}}-S_{Mod_{i}}\right)^{2}\right]} \times 1000,
\end{equation}
where $n$ is the number of wavelengths, $m$ is the number of fitting parameters and $\mathrm{N}=\operatorname{Cos}(2 \Psi), \mathrm{C}=\operatorname{Sin}(2 \Psi) \operatorname{Cos}(\Delta), \mathrm{S}=\operatorname{Sin}(2 \Psi) \operatorname{Sin}(\Delta)$. The subscripts $Exp$ and $Mod$ denote the experimental data and the modeled data, respectively.

\subsection{Spectrophotometry}

A JASCO V-570 UV/Vis/NIR spectrophotometer was used to measure the optical transmission and specular reflectance of MAPbI$_3$ thin films deposited on quartz. The data was used to provide an estimate for the absorption coefficient using the relation
\begin{equation}
\alpha(E)=\frac{1}{d} \ln \left(\frac{(1-R(E))^{2}}{T(E)}\right),
\label {Absco}
\end{equation}
where $d$ is the film thickness, $T(E)$ is the transmission and $R(E)$ is the reflectance.

\begin{acknowledgement}

The authors thank the Engineering and Physical Sciences Research Council (EPSRC) for funding the research.

\end{acknowledgement}




\bibliography{references.bib}

\providecommand{\latin}[1]{#1}
\makeatletter
\providecommand{\doi}
  {\begingroup\let\do\@makeother\dospecials
  \catcode`\{=1 \catcode`\}=2 \doi@aux}
\providecommand{\doi@aux}[1]{\endgroup\texttt{#1}}
\makeatother
\providecommand*\mcitethebibliography{\thebibliography}
\csname @ifundefined\endcsname{endmcitethebibliography}
  {\let\endmcitethebibliography\endthebibliography}{}
\begin{mcitethebibliography}{61}
\providecommand*\natexlab[1]{#1}
\providecommand*\mciteSetBstSublistMode[1]{}
\providecommand*\mciteSetBstMaxWidthForm[2]{}
\providecommand*\mciteBstWouldAddEndPuncttrue
  {\def\EndOfBibitem{\unskip.}}
\providecommand*\mciteBstWouldAddEndPunctfalse
  {\let\EndOfBibitem\relax}
\providecommand*\mciteSetBstMidEndSepPunct[3]{}
\providecommand*\mciteSetBstSublistLabelBeginEnd[3]{}
\providecommand*\EndOfBibitem{}
\mciteSetBstSublistMode{f}
\mciteSetBstMaxWidthForm{subitem}{(\alph{mcitesubitemcount})}
\mciteSetBstSublistLabelBeginEnd
  {\mcitemaxwidthsubitemform\space}
  {\relax}
  {\relax}

\bibitem[{NREL}(2019)]{NREL2019}
{NREL}, {NREL Efficiency Chart}. 2019;
  \url{https://www.nrel.gov/pv/cell-efficiency.html}\relax
\mciteBstWouldAddEndPuncttrue
\mciteSetBstMidEndSepPunct{\mcitedefaultmidpunct}
{\mcitedefaultendpunct}{\mcitedefaultseppunct}\relax
\EndOfBibitem
\bibitem[Correa-Baena \latin{et~al.}(2017)Correa-Baena, Saliba, Buonassisi,
  Gr{\"{a}}tzel, Abate, Tress, and Hagfeldt]{Correa-Baena2017}
Correa-Baena,~J.~P.; Saliba,~M.; Buonassisi,~T.; Gr{\"{a}}tzel,~M.; Abate,~A.;
  Tress,~W.; Hagfeldt,~A. \emph{Science} \textbf{2017}, \emph{358},
  739--744\relax
\mciteBstWouldAddEndPuncttrue
\mciteSetBstMidEndSepPunct{\mcitedefaultmidpunct}
{\mcitedefaultendpunct}{\mcitedefaultseppunct}\relax
\EndOfBibitem
\bibitem[Sha \latin{et~al.}(2015)Sha, Ren, Chen, and Choy]{Sha2015}
Sha,~W.~E.; Ren,~X.; Chen,~L.; Choy,~W.~C. \emph{Applied Physics Letters}
  \textbf{2015}, \emph{106}, 221104\relax
\mciteBstWouldAddEndPuncttrue
\mciteSetBstMidEndSepPunct{\mcitedefaultmidpunct}
{\mcitedefaultendpunct}{\mcitedefaultseppunct}\relax
\EndOfBibitem
\bibitem[Polman \latin{et~al.}(2016)Polman, Knight, Garnett, Ehrler, and
  Sinke]{Polman2016}
Polman,~A.; Knight,~M.; Garnett,~E.~C.; Ehrler,~B.; Sinke,~W.~C. \emph{Science}
  \textbf{2016}, \emph{352}, aad4424--aad4424\relax
\mciteBstWouldAddEndPuncttrue
\mciteSetBstMidEndSepPunct{\mcitedefaultmidpunct}
{\mcitedefaultendpunct}{\mcitedefaultseppunct}\relax
\EndOfBibitem
\bibitem[Braly \latin{et~al.}(2018)Braly, Dequilettes, Pazos-Out{\'{o}}n,
  Burke, Ziffer, Ginger, and Hillhouse]{Braly2018}
Braly,~I.~L.; Dequilettes,~D.~W.; Pazos-Out{\'{o}}n,~L.~M.; Burke,~S.;
  Ziffer,~M.~E.; Ginger,~D.~S.; Hillhouse,~H.~W. \emph{Nature Photonics}
  \textbf{2018}, \emph{12}, 355--361\relax
\mciteBstWouldAddEndPuncttrue
\mciteSetBstMidEndSepPunct{\mcitedefaultmidpunct}
{\mcitedefaultendpunct}{\mcitedefaultseppunct}\relax
\EndOfBibitem
\bibitem[Paetzold \latin{et~al.}(2017)Paetzold, Jaysankar, Gehlhaar, Ahlswede,
  Paetel, Qiu, Bastos, Rakocevic, Richards, Aernouts, Powalla, and
  Poortmans]{Paetzold2017}
Paetzold,~U. W.~W.; Jaysankar,~M.; Gehlhaar,~R.; Ahlswede,~E.; Paetel,~S.;
  Qiu,~W.; Bastos,~J.~P.; Rakocevic,~L.; Richards,~B.~S.; Aernouts,~T.;
  Powalla,~M.; Poortmans,~J. \emph{J. Mater. Chem. A} \textbf{2017}, \emph{5},
  9897--9906\relax
\mciteBstWouldAddEndPuncttrue
\mciteSetBstMidEndSepPunct{\mcitedefaultmidpunct}
{\mcitedefaultendpunct}{\mcitedefaultseppunct}\relax
\EndOfBibitem
\bibitem[Albrecht \latin{et~al.}(2016)Albrecht, Saliba, Correa~Baena, Lang,
  Kegelmann, Mews, Steier, Abate, Rappich, Korte, Schlatmann, Nazeeruddin,
  Hagfeldt, Gr{\"{a}}tzel, and Rech]{Albrecht2016}
Albrecht,~S.; Saliba,~M.; Correa~Baena,~J.~P.; Lang,~F.; Kegelmann,~L.;
  Mews,~M.; Steier,~L.; Abate,~A.; Rappich,~J.; Korte,~L.; Schlatmann,~R.;
  Nazeeruddin,~M.~K.; Hagfeldt,~A.; Gr{\"{a}}tzel,~M.; Rech,~B. \emph{Energy
  Environ. Sci.} \textbf{2016}, \emph{9}, 81--88\relax
\mciteBstWouldAddEndPuncttrue
\mciteSetBstMidEndSepPunct{\mcitedefaultmidpunct}
{\mcitedefaultendpunct}{\mcitedefaultseppunct}\relax
\EndOfBibitem
\bibitem[Jiang \latin{et~al.}(2016)Jiang, Almansouri, Huang, Young, Li, Peng,
  Hou, Spiccia, Bach, Cheng, Green, and Ho-Baillie]{Jiang2016}
Jiang,~Y.; Almansouri,~I.; Huang,~S.; Young,~T.; Li,~Y.; Peng,~Y.; Hou,~Q.;
  Spiccia,~L.; Bach,~U.; Cheng,~Y.-B.; Green,~M.; Ho-Baillie,~A. \emph{J.
  Mater. Chem. C} \textbf{2016}, \emph{4}, 5679--5689\relax
\mciteBstWouldAddEndPuncttrue
\mciteSetBstMidEndSepPunct{\mcitedefaultmidpunct}
{\mcitedefaultendpunct}{\mcitedefaultseppunct}\relax
\EndOfBibitem
\bibitem[Mantilla-Perez \latin{et~al.}(2017)Mantilla-Perez, Feurer,
  Correa-Baena, Liu, Colodrero, Toudert, Saliba, Buecheler, Hagfeldt, Tiwari,
  and Martorell]{Mantilla-Perez2017}
Mantilla-Perez,~P.; Feurer,~T.; Correa-Baena,~J.~P.; Liu,~Q.; Colodrero,~S.;
  Toudert,~J.; Saliba,~M.; Buecheler,~S.; Hagfeldt,~A.; Tiwari,~A.~N.;
  Martorell,~J. \emph{ACS Photonics} \textbf{2017}, \emph{4}, 861--867\relax
\mciteBstWouldAddEndPuncttrue
\mciteSetBstMidEndSepPunct{\mcitedefaultmidpunct}
{\mcitedefaultendpunct}{\mcitedefaultseppunct}\relax
\EndOfBibitem
\bibitem[Shirayama \latin{et~al.}(2016)Shirayama, Kadowaki, Miyadera, Sugita,
  Tamakoshi, Kato, Fujiseki, Murata, Hara, Murakami, Fujimoto, Chikamatsu, and
  Fujiwara]{Shirayama2016}
Shirayama,~M.; Kadowaki,~H.; Miyadera,~T.; Sugita,~T.; Tamakoshi,~M.; Kato,~M.;
  Fujiseki,~T.; Murata,~D.; Hara,~S.; Murakami,~T.~N.; Fujimoto,~S.;
  Chikamatsu,~M.; Fujiwara,~H. \emph{Physical Review Applied} \textbf{2016},
  \emph{5}, 014012\relax
\mciteBstWouldAddEndPuncttrue
\mciteSetBstMidEndSepPunct{\mcitedefaultmidpunct}
{\mcitedefaultendpunct}{\mcitedefaultseppunct}\relax
\EndOfBibitem
\bibitem[De~Wolf \latin{et~al.}(2014)De~Wolf, Holovsky, Moon, L{\"{o}}per,
  Niesen, Ledinsky, Haug, Yum, and Ballif]{DeWolf2014}
De~Wolf,~S.; Holovsky,~J.; Moon,~S.~J.; L{\"{o}}per,~P.; Niesen,~B.;
  Ledinsky,~M.; Haug,~F.~J.; Yum,~J.~H.; Ballif,~C. \emph{Journal of Physical
  Chemistry Letters} \textbf{2014}, \emph{5}, 1035--1039\relax
\mciteBstWouldAddEndPuncttrue
\mciteSetBstMidEndSepPunct{\mcitedefaultmidpunct}
{\mcitedefaultendpunct}{\mcitedefaultseppunct}\relax
\EndOfBibitem
\bibitem[Xing \latin{et~al.}(2013)Xing, Mathews, Sun, Lim, Lam, Gr{\"{a}}tzel,
  Mhaisalkar, and Sum]{Xing2013}
Xing,~G.; Mathews,~N.; Sun,~S.; Lim,~S.~S.; Lam,~Y.~M.; Gr{\"{a}}tzel,~M.;
  Mhaisalkar,~S.; Sum,~T.~C. \emph{Science} \textbf{2013}, \emph{342}, 344 LP
  -- 347\relax
\mciteBstWouldAddEndPuncttrue
\mciteSetBstMidEndSepPunct{\mcitedefaultmidpunct}
{\mcitedefaultendpunct}{\mcitedefaultseppunct}\relax
\EndOfBibitem
\bibitem[Sun \latin{et~al.}(2014)Sun, Salim, Mathews, Duchamp, Boothroyd, Xing,
  Sum, and Lam]{Sun2014}
Sun,~S.; Salim,~T.; Mathews,~N.; Duchamp,~M.; Boothroyd,~C.; Xing,~G.;
  Sum,~T.~C.; Lam,~Y.~M. \emph{Energy {\&} Environmental Science}
  \textbf{2014}, \emph{7}, 399\relax
\mciteBstWouldAddEndPuncttrue
\mciteSetBstMidEndSepPunct{\mcitedefaultmidpunct}
{\mcitedefaultendpunct}{\mcitedefaultseppunct}\relax
\EndOfBibitem
\bibitem[Miyata \latin{et~al.}(2015)Miyata, Mitioglu, Plochocka, Portugall,
  Wang, Stranks, Snaith, and Nicholas]{Miyata2015}
Miyata,~A.; Mitioglu,~A.; Plochocka,~P.; Portugall,~O.; Wang,~J. T.-W.;
  Stranks,~S.~D.; Snaith,~H.~J.; Nicholas,~R.~J. \emph{Nature Physics}
  \textbf{2015}, \emph{11}, 582--587\relax
\mciteBstWouldAddEndPuncttrue
\mciteSetBstMidEndSepPunct{\mcitedefaultmidpunct}
{\mcitedefaultendpunct}{\mcitedefaultseppunct}\relax
\EndOfBibitem
\bibitem[Even \latin{et~al.}(2014)Even, Pedesseau, and Katan]{Even2014}
Even,~J.; Pedesseau,~L.; Katan,~C. \emph{Journal of Physical Chemistry C}
  \textbf{2014}, \emph{118}, 11566--11572\relax
\mciteBstWouldAddEndPuncttrue
\mciteSetBstMidEndSepPunct{\mcitedefaultmidpunct}
{\mcitedefaultendpunct}{\mcitedefaultseppunct}\relax
\EndOfBibitem
\bibitem[Yang \latin{et~al.}(2015)Yang, Yang, Li, Crisp, Zhu, and
  Beard]{yang2015}
Yang,~Y.; Yang,~M.; Li,~Z.; Crisp,~R.; Zhu,~K.; Beard,~M.~C. \emph{Journal of
  Physical Chemistry Letters} \textbf{2015}, \emph{6}, 4688--4692\relax
\mciteBstWouldAddEndPuncttrue
\mciteSetBstMidEndSepPunct{\mcitedefaultmidpunct}
{\mcitedefaultendpunct}{\mcitedefaultseppunct}\relax
\EndOfBibitem
\bibitem[Piana \latin{et~al.}(2019)Piana, Bailey, and Lagoudakis]{Piana2019}
Piana,~G.~M.; Bailey,~C.~G.; Lagoudakis,~P.~G. \emph{The Journal of Physical
  Chemistry C} \textbf{2019}, \emph{37}, acs.jpcc.9b06712\relax
\mciteBstWouldAddEndPuncttrue
\mciteSetBstMidEndSepPunct{\mcitedefaultmidpunct}
{\mcitedefaultendpunct}{\mcitedefaultseppunct}\relax
\EndOfBibitem
\bibitem[Giorgi \latin{et~al.}(2013)Giorgi, Fujisawa, Segawa, and
  Yamashita]{Giorgi2013}
Giorgi,~G.; Fujisawa,~J.~I.; Segawa,~H.; Yamashita,~K. \emph{Journal of
  Physical Chemistry Letters} \textbf{2013}, \emph{4}, 4213--4216\relax
\mciteBstWouldAddEndPuncttrue
\mciteSetBstMidEndSepPunct{\mcitedefaultmidpunct}
{\mcitedefaultendpunct}{\mcitedefaultseppunct}\relax
\EndOfBibitem
\bibitem[Ziffer \latin{et~al.}(2016)Ziffer, Mohammed, and Ginger]{Ziffer2016}
Ziffer,~M.~E.; Mohammed,~J.~C.; Ginger,~D.~S. \emph{ACS Photonics}
  \textbf{2016}, \emph{3}, 1060--1068\relax
\mciteBstWouldAddEndPuncttrue
\mciteSetBstMidEndSepPunct{\mcitedefaultmidpunct}
{\mcitedefaultendpunct}{\mcitedefaultseppunct}\relax
\EndOfBibitem
\bibitem[Li \latin{et~al.}(2018)Li, Moon, Gharajeh, Haroldson, Hawkins, Hu,
  Zakhidov, and Gu]{Li2018a}
Li,~Z.; Moon,~J.; Gharajeh,~A.; Haroldson,~R.; Hawkins,~R.; Hu,~W.;
  Zakhidov,~A.; Gu,~Q. \emph{ACS Nano} \textbf{2018}, \emph{12},
  10968--10976\relax
\mciteBstWouldAddEndPuncttrue
\mciteSetBstMidEndSepPunct{\mcitedefaultmidpunct}
{\mcitedefaultendpunct}{\mcitedefaultseppunct}\relax
\EndOfBibitem
\bibitem[Wang \latin{et~al.}(2018)Wang, Liu, Li, Zhang, Zhang, Xiao, and
  Song]{Wang2018}
Wang,~S.; Liu,~Y.; Li,~G.; Zhang,~J.; Zhang,~N.; Xiao,~S.; Song,~Q.
  \emph{Advanced Optical Materials} \textbf{2018}, \emph{6}, 1701266\relax
\mciteBstWouldAddEndPuncttrue
\mciteSetBstMidEndSepPunct{\mcitedefaultmidpunct}
{\mcitedefaultendpunct}{\mcitedefaultseppunct}\relax
\EndOfBibitem
\bibitem[Wang \latin{et~al.}(2016)Wang, Sun, Li, Gu, Xiao, and Song]{Wang2016}
Wang,~K.; Sun,~W.; Li,~J.; Gu,~Z.; Xiao,~S.; Song,~Q. \emph{ACS Photonics}
  \textbf{2016}, \emph{3}, 1125--1130\relax
\mciteBstWouldAddEndPuncttrue
\mciteSetBstMidEndSepPunct{\mcitedefaultmidpunct}
{\mcitedefaultendpunct}{\mcitedefaultseppunct}\relax
\EndOfBibitem
\bibitem[Stylianakis \latin{et~al.}(2019)Stylianakis, Maksudov,
  Panagiotopoulos, Kakavelakis, and Petridis]{Stylianakis2019}
Stylianakis,~M.~M.; Maksudov,~T.; Panagiotopoulos,~A.; Kakavelakis,~G.;
  Petridis,~K. \emph{Materials} \textbf{2019}, \emph{16}, 1--28\relax
\mciteBstWouldAddEndPuncttrue
\mciteSetBstMidEndSepPunct{\mcitedefaultmidpunct}
{\mcitedefaultendpunct}{\mcitedefaultseppunct}\relax
\EndOfBibitem
\bibitem[Liao \latin{et~al.}(2015)Liao, Hu, Zhang, Wang, Yao, and Fu]{Liao2015}
Liao,~Q.; Hu,~K.; Zhang,~H.; Wang,~X.; Yao,~J.; Fu,~H. \emph{Advanced
  Materials} \textbf{2015}, \emph{27}, 3405--3410\relax
\mciteBstWouldAddEndPuncttrue
\mciteSetBstMidEndSepPunct{\mcitedefaultmidpunct}
{\mcitedefaultendpunct}{\mcitedefaultseppunct}\relax
\EndOfBibitem
\bibitem[Zhu \latin{et~al.}(2015)Zhu, Fu, Meng, Wu, Gong, Ding, Gustafsson,
  Trinh, Jin, and Zhu]{Zhu2015}
Zhu,~H.; Fu,~Y.; Meng,~F.; Wu,~X.; Gong,~Z.; Ding,~Q.; Gustafsson,~M.~V.;
  Trinh,~M.~T.; Jin,~S.; Zhu,~X.~Y. \emph{Nature Materials} \textbf{2015},
  \emph{14}, 636--642\relax
\mciteBstWouldAddEndPuncttrue
\mciteSetBstMidEndSepPunct{\mcitedefaultmidpunct}
{\mcitedefaultendpunct}{\mcitedefaultseppunct}\relax
\EndOfBibitem
\bibitem[Tan \latin{et~al.}(2014)Tan, Moghaddam, Lai, Docampo, Higler,
  Deschler, Price, Sadhanala, Pazos, Credgington, Hanusch, Bein, Snaith, and
  Friend]{Tan2014}
Tan,~Z.~K.; Moghaddam,~R.~S.; Lai,~M.~L.; Docampo,~P.; Higler,~R.;
  Deschler,~F.; Price,~M.; Sadhanala,~A.; Pazos,~L.~M.; Credgington,~D.;
  Hanusch,~F.; Bein,~T.; Snaith,~H.~J.; Friend,~R.~H. \emph{Nature
  Nanotechnology} \textbf{2014}, \emph{9}, 687--692\relax
\mciteBstWouldAddEndPuncttrue
\mciteSetBstMidEndSepPunct{\mcitedefaultmidpunct}
{\mcitedefaultendpunct}{\mcitedefaultseppunct}\relax
\EndOfBibitem
\bibitem[Cho \latin{et~al.}(2015)Cho, Jeong, Park, Kim, Wolf, Lee, Heo,
  Sadhanala, Myoung, Yoo, Im, Friend, and Lee]{Cho2015}
Cho,~H.; Jeong,~S.~H.; Park,~M.~H.; Kim,~Y.~H.; Wolf,~C.; Lee,~C.~L.;
  Heo,~J.~H.; Sadhanala,~A.; Myoung,~N.~S.; Yoo,~S.; Im,~S.~H.; Friend,~R.~H.;
  Lee,~T.~W. \emph{Science} \textbf{2015}, \emph{350}, 1222--1225\relax
\mciteBstWouldAddEndPuncttrue
\mciteSetBstMidEndSepPunct{\mcitedefaultmidpunct}
{\mcitedefaultendpunct}{\mcitedefaultseppunct}\relax
\EndOfBibitem
\bibitem[Ling \latin{et~al.}(2016)Ling, Yuan, Tian, Wang, Wang, Xin, Hanson,
  Ma, and Gao]{Ling2016}
Ling,~Y.; Yuan,~Z.; Tian,~Y.; Wang,~X.; Wang,~J.~C.; Xin,~Y.; Hanson,~K.;
  Ma,~B.; Gao,~H. \emph{Advanced Materials} \textbf{2016}, \emph{28},
  305--311\relax
\mciteBstWouldAddEndPuncttrue
\mciteSetBstMidEndSepPunct{\mcitedefaultmidpunct}
{\mcitedefaultendpunct}{\mcitedefaultseppunct}\relax
\EndOfBibitem
\bibitem[Liang \latin{et~al.}(2016)Liang, Zhang, Guo, Gan, Lin, Fan, and
  Liu]{Liang2016}
Liang,~J.; Zhang,~Y.; Guo,~X.; Gan,~Z.; Lin,~J.; Fan,~Y.; Liu,~X. \emph{RSC
  Advances} \textbf{2016}, \emph{6}, 71070--71075\relax
\mciteBstWouldAddEndPuncttrue
\mciteSetBstMidEndSepPunct{\mcitedefaultmidpunct}
{\mcitedefaultendpunct}{\mcitedefaultseppunct}\relax
\EndOfBibitem
\bibitem[Saidaminov \latin{et~al.}(2015)Saidaminov, Adinolfi, Comin, Abdelhady,
  Peng, Dursun, Yuan, Hoogland, Sargent, and Bakr]{Saidaminov2015}
Saidaminov,~M.~I.; Adinolfi,~V.; Comin,~R.; Abdelhady,~A.~L.; Peng,~W.;
  Dursun,~I.; Yuan,~M.; Hoogland,~S.; Sargent,~E.~H.; Bakr,~O.~M. \emph{Nature
  Communications} \textbf{2015}, \emph{6}, 8724\relax
\mciteBstWouldAddEndPuncttrue
\mciteSetBstMidEndSepPunct{\mcitedefaultmidpunct}
{\mcitedefaultendpunct}{\mcitedefaultseppunct}\relax
\EndOfBibitem
\bibitem[Fang \latin{et~al.}(2015)Fang, Dong, Shao, Yuan, and Huang]{Fang2015a}
Fang,~Y.; Dong,~Q.; Shao,~Y.; Yuan,~Y.; Huang,~J. \emph{Nature Photonics}
  \textbf{2015}, \emph{9}, 679--686\relax
\mciteBstWouldAddEndPuncttrue
\mciteSetBstMidEndSepPunct{\mcitedefaultmidpunct}
{\mcitedefaultendpunct}{\mcitedefaultseppunct}\relax
\EndOfBibitem
\bibitem[Bao \latin{et~al.}(2017)Bao, Chen, Fang, Wei, Deng, Xiao, Li, and
  Huang]{Bao2017}
Bao,~C.; Chen,~Z.; Fang,~Y.; Wei,~H.; Deng,~Y.; Xiao,~X.; Li,~L.; Huang,~J.
  \emph{Advanced Materials} \textbf{2017}, \emph{29}, 1703209\relax
\mciteBstWouldAddEndPuncttrue
\mciteSetBstMidEndSepPunct{\mcitedefaultmidpunct}
{\mcitedefaultendpunct}{\mcitedefaultseppunct}\relax
\EndOfBibitem
\bibitem[Ball \latin{et~al.}(2015)Ball, Stranks, H{\"{o}}rantner,
  H{\"{u}}ttner, Zhang, Crossland, Ramirez, Riede, Johnston, Friend, and
  Snaith]{Ball2015}
Ball,~J.~M.; Stranks,~S.~D.; H{\"{o}}rantner,~M.~T.; H{\"{u}}ttner,~S.;
  Zhang,~W.; Crossland,~E.~J.; Ramirez,~I.; Riede,~M.; Johnston,~M.~B.;
  Friend,~R.~H.; Snaith,~H.~J. \emph{Energy and Environmental Science}
  \textbf{2015}, \emph{8}, 602--609\relax
\mciteBstWouldAddEndPuncttrue
\mciteSetBstMidEndSepPunct{\mcitedefaultmidpunct}
{\mcitedefaultendpunct}{\mcitedefaultseppunct}\relax
\EndOfBibitem
\bibitem[Ziang \latin{et~al.}(2015)Ziang, Shifeng, Laixiang, Shuping, Wei, Yu,
  Li, Zhijian, Shufeng, Honglin, Minghui, and Qin]{Ziang2015}
Ziang,~X.; Shifeng,~L.; Laixiang,~Q.; Shuping,~P.; Wei,~W.; Yu,~Y.; Li,~Y.;
  Zhijian,~C.; Shufeng,~W.; Honglin,~D.; Minghui,~Y.; Qin,~G.~G. \emph{Optical
  Materials Express} \textbf{2015}, \emph{5}, 29\relax
\mciteBstWouldAddEndPuncttrue
\mciteSetBstMidEndSepPunct{\mcitedefaultmidpunct}
{\mcitedefaultendpunct}{\mcitedefaultseppunct}\relax
\EndOfBibitem
\bibitem[Jiang \latin{et~al.}(2015)Jiang, Green, Sheng, and
  Ho-Baillie]{Jiang2015}
Jiang,~Y.; Green,~M.~A.; Sheng,~R.; Ho-Baillie,~A. \emph{Solar Energy Materials
  and Solar Cells} \textbf{2015}, \emph{137}, 253--257\relax
\mciteBstWouldAddEndPuncttrue
\mciteSetBstMidEndSepPunct{\mcitedefaultmidpunct}
{\mcitedefaultendpunct}{\mcitedefaultseppunct}\relax
\EndOfBibitem
\bibitem[Marronnier \latin{et~al.}(2018)Marronnier, Lee, Lee, Kim, Eypert,
  Gaston, Roma, Tondelier, Geffroy, and Bonnassieux]{Marronnier2018}
Marronnier,~A.; Lee,~H.; Lee,~H.; Kim,~M.; Eypert,~C.; Gaston,~J.~P.; Roma,~G.;
  Tondelier,~D.; Geffroy,~B.; Bonnassieux,~Y. \emph{Solar Energy Materials and
  Solar Cells} \textbf{2018}, \emph{178}, 179--185\relax
\mciteBstWouldAddEndPuncttrue
\mciteSetBstMidEndSepPunct{\mcitedefaultmidpunct}
{\mcitedefaultendpunct}{\mcitedefaultseppunct}\relax
\EndOfBibitem
\bibitem[Wang \latin{et~al.}(2019)Wang, Gong, Shan, Zhang, Xu, Dai, Wang, Wang,
  Fang, and Zhang]{Wang2019}
Wang,~X.; Gong,~J.; Shan,~X.; Zhang,~M.; Xu,~Z.; Dai,~R.; Wang,~Z.; Wang,~S.;
  Fang,~X.; Zhang,~Z. \emph{Journal of Physical Chemistry C} \textbf{2019},
  \emph{123}, 1362--1369\relax
\mciteBstWouldAddEndPuncttrue
\mciteSetBstMidEndSepPunct{\mcitedefaultmidpunct}
{\mcitedefaultendpunct}{\mcitedefaultseppunct}\relax
\EndOfBibitem
\bibitem[Lin \latin{et~al.}(2015)Lin, Armin, Nagiri, Burn, and
  Meredith]{Lin2015}
Lin,~Q.; Armin,~A.; Nagiri,~R. C.~R.; Burn,~P.~L.; Meredith,~P. \emph{Nature
  Photonics} \textbf{2015}, \emph{9}, 106--112\relax
\mciteBstWouldAddEndPuncttrue
\mciteSetBstMidEndSepPunct{\mcitedefaultmidpunct}
{\mcitedefaultendpunct}{\mcitedefaultseppunct}\relax
\EndOfBibitem
\bibitem[Leguy \latin{et~al.}(2015)Leguy, Hu, Campoy-Quiles, Alonso, Weber,
  Azarhoosh, Van~Schilfgaarde, Weller, Bein, Nelson, Docampo, and
  Barnes]{Leguy2015}
Leguy,~A.~M.; Hu,~Y.; Campoy-Quiles,~M.; Alonso,~M.~I.; Weber,~O.~J.;
  Azarhoosh,~P.; Van~Schilfgaarde,~M.; Weller,~M.~T.; Bein,~T.; Nelson,~J.;
  Docampo,~P.; Barnes,~P.~R. \emph{Chemistry of Materials} \textbf{2015},
  \emph{27}, 3397--3407\relax
\mciteBstWouldAddEndPuncttrue
\mciteSetBstMidEndSepPunct{\mcitedefaultmidpunct}
{\mcitedefaultendpunct}{\mcitedefaultseppunct}\relax
\EndOfBibitem
\bibitem[Leguy \latin{et~al.}(2016)Leguy, Azarhoosh, Alonso, Campoy-Quiles,
  Weber, Yao, Bryant, Weller, Nelson, Walsh, Van~Schilfgaarde, and
  Barnes]{Leguy2016a}
Leguy,~A.~M.; Azarhoosh,~P.; Alonso,~M.~I.; Campoy-Quiles,~M.; Weber,~O.~J.;
  Yao,~J.; Bryant,~D.; Weller,~M.~T.; Nelson,~J.; Walsh,~A.;
  Van~Schilfgaarde,~M.; Barnes,~P.~R. \emph{Nanoscale} \textbf{2016}, \emph{8},
  6317--6327\relax
\mciteBstWouldAddEndPuncttrue
\mciteSetBstMidEndSepPunct{\mcitedefaultmidpunct}
{\mcitedefaultendpunct}{\mcitedefaultseppunct}\relax
\EndOfBibitem
\bibitem[L{\"{o}}per \latin{et~al.}(2015)L{\"{o}}per, Stuckelberger, Niesen,
  Werner, Filipi{\v{c}}, Moon, Yum, Topi{\v{c}}, De~Wolf, and
  Ballif]{Loper2015}
L{\"{o}}per,~P.; Stuckelberger,~M.; Niesen,~B.; Werner,~J.; Filipi{\v{c}},~M.;
  Moon,~S.~J.; Yum,~J.~H.; Topi{\v{c}},~M.; De~Wolf,~S.; Ballif,~C.
  \emph{Journal of Physical Chemistry Letters} \textbf{2015}, \emph{6},
  66--71\relax
\mciteBstWouldAddEndPuncttrue
\mciteSetBstMidEndSepPunct{\mcitedefaultmidpunct}
{\mcitedefaultendpunct}{\mcitedefaultseppunct}\relax
\EndOfBibitem
\bibitem[Jiang \latin{et~al.}(2016)Jiang, Soufiani, Gentle, Huang, Ho-Baillie,
  and Green]{Jiang2016a}
Jiang,~Y.; Soufiani,~A.~M.; Gentle,~A.; Huang,~F.; Ho-Baillie,~A.; Green,~M.~A.
  \emph{Applied Physics Letters} \textbf{2016}, \emph{108}, 61905\relax
\mciteBstWouldAddEndPuncttrue
\mciteSetBstMidEndSepPunct{\mcitedefaultmidpunct}
{\mcitedefaultendpunct}{\mcitedefaultseppunct}\relax
\EndOfBibitem
\bibitem[van Eerden \latin{et~al.}(2017)van Eerden, Jaysankar, Hadipour,
  Merckx, Schermer, Aernouts, Poortmans, and Paetzold]{VanEerden2017}
van Eerden,~M.; Jaysankar,~M.; Hadipour,~A.; Merckx,~T.; Schermer,~J.~J.;
  Aernouts,~T.; Poortmans,~J.; Paetzold,~U.~W. \emph{Advanced Optical
  Materials} \textbf{2017}, \emph{5}, 1700151\relax
\mciteBstWouldAddEndPuncttrue
\mciteSetBstMidEndSepPunct{\mcitedefaultmidpunct}
{\mcitedefaultendpunct}{\mcitedefaultseppunct}\relax
\EndOfBibitem
\bibitem[Demchenko \latin{et~al.}(2016)Demchenko, Izyumskaya, Feneberg,
  Avrutin, {{\"{O}}zg{\"{u}}r}, Goldhahn, and Morko{\c{c}}]{Demchenko2016}
Demchenko,~D.~O.; Izyumskaya,~N.; Feneberg,~M.; Avrutin,~V.;
  {{\"{O}}zg{\"{u}}r},; Goldhahn,~R.; Morko{\c{c}},~H. \emph{Physical Review B}
  \textbf{2016}, \emph{94}, 75206\relax
\mciteBstWouldAddEndPuncttrue
\mciteSetBstMidEndSepPunct{\mcitedefaultmidpunct}
{\mcitedefaultendpunct}{\mcitedefaultseppunct}\relax
\EndOfBibitem
\bibitem[Ndione \latin{et~al.}(2016)Ndione, Li, and Zhu]{Ndione2016}
Ndione,~P.~F.; Li,~Z.; Zhu,~K. \emph{Journal of Materials Chemistry C}
  \textbf{2016}, \emph{4}, 7775--7782\relax
\mciteBstWouldAddEndPuncttrue
\mciteSetBstMidEndSepPunct{\mcitedefaultmidpunct}
{\mcitedefaultendpunct}{\mcitedefaultseppunct}\relax
\EndOfBibitem
\bibitem[Guerra \latin{et~al.}(2017)Guerra, Tejada, Korte, Kegelmann,
  T{\"{o}}fflinger, Albrecht, Rech, and Weing{\"{a}}rtner]{Guerra2017}
Guerra,~J.~A.; Tejada,~A.; Korte,~L.; Kegelmann,~L.; T{\"{o}}fflinger,~J.~A.;
  Albrecht,~S.; Rech,~B.; Weing{\"{a}}rtner,~R. \emph{Journal of Applied
  Physics} \textbf{2017}, \emph{121}, 173104\relax
\mciteBstWouldAddEndPuncttrue
\mciteSetBstMidEndSepPunct{\mcitedefaultmidpunct}
{\mcitedefaultendpunct}{\mcitedefaultseppunct}\relax
\EndOfBibitem
\bibitem[Brittman and Garnett(2016)Brittman, and Garnett]{Brittman2016}
Brittman,~S.; Garnett,~E.~C. \emph{Journal of Physical Chemistry C}
  \textbf{2016}, \emph{120}, 616--620\relax
\mciteBstWouldAddEndPuncttrue
\mciteSetBstMidEndSepPunct{\mcitedefaultmidpunct}
{\mcitedefaultendpunct}{\mcitedefaultseppunct}\relax
\EndOfBibitem
\bibitem[Chen \latin{et~al.}(2019)Chen, Wang, Song, Li, Xu, Zeng, and
  Sun]{Chen2019}
Chen,~X.; Wang,~Y.; Song,~J.; Li,~X.; Xu,~J.; Zeng,~H.; Sun,~H. \emph{Journal
  of Physical Chemistry C} \textbf{2019}, \emph{123}, 10564--10570\relax
\mciteBstWouldAddEndPuncttrue
\mciteSetBstMidEndSepPunct{\mcitedefaultmidpunct}
{\mcitedefaultendpunct}{\mcitedefaultseppunct}\relax
\EndOfBibitem
\bibitem[Shirai(2017)]{Shirai2017}
Shirai,~H. \emph{Ellipsometry - Principles and Techniques for Materials
  Characterization}; 2017\relax
\mciteBstWouldAddEndPuncttrue
\mciteSetBstMidEndSepPunct{\mcitedefaultmidpunct}
{\mcitedefaultendpunct}{\mcitedefaultseppunct}\relax
\EndOfBibitem
\bibitem[Zhao \latin{et~al.}(2018)Zhao, Shi, Dai, and Lian]{Zhao2018}
Zhao,~M.; Shi,~Y.; Dai,~J.; Lian,~J. \emph{Journal of Materials Chemistry C}
  \textbf{2018}, \emph{6}, 10450--10455\relax
\mciteBstWouldAddEndPuncttrue
\mciteSetBstMidEndSepPunct{\mcitedefaultmidpunct}
{\mcitedefaultendpunct}{\mcitedefaultseppunct}\relax
\EndOfBibitem
\bibitem[Alias \latin{et~al.}(2016)Alias, Dursun, Saidaminov, Diallo, Mishra,
  Ng, Bakr, and Ooi]{Alias2016}
Alias,~M.~S.; Dursun,~I.; Saidaminov,~M.~I.; Diallo,~E.~M.; Mishra,~P.;
  Ng,~T.~K.; Bakr,~O.~M.; Ooi,~B.~S. \emph{Optics Express} \textbf{2016},
  \emph{24}, 16586\relax
\mciteBstWouldAddEndPuncttrue
\mciteSetBstMidEndSepPunct{\mcitedefaultmidpunct}
{\mcitedefaultendpunct}{\mcitedefaultseppunct}\relax
\EndOfBibitem
\bibitem[Werner \latin{et~al.}(2018)Werner, Nogay, Sahli, Yang,
  Br{\"{a}}uninger, Christmann, Walter, Kamino, Fiala, L{\"{o}}per, Nicolay,
  Jeangros, Niesen, and Ballif]{Werner2018}
Werner,~J.; Nogay,~G.; Sahli,~F.; Yang,~T. C.~J.; Br{\"{a}}uninger,~M.;
  Christmann,~G.; Walter,~A.; Kamino,~B.~A.; Fiala,~P.; L{\"{o}}per,~P.;
  Nicolay,~S.; Jeangros,~Q.; Niesen,~B.; Ballif,~C. \emph{ACS Energy Letters}
  \textbf{2018}, \emph{3}, 742--747\relax
\mciteBstWouldAddEndPuncttrue
\mciteSetBstMidEndSepPunct{\mcitedefaultmidpunct}
{\mcitedefaultendpunct}{\mcitedefaultseppunct}\relax
\EndOfBibitem
\bibitem[Whitcher \latin{et~al.}(2018)Whitcher, Zhu, Chi, Hu, Zhao, Asmara, Yu,
  Breese, Castro~Neto, Lam, Wee, Chia, and Rusydi]{Whitcher2018}
Whitcher,~T.~J.; Zhu,~J.~X.; Chi,~X.; Hu,~H.; Zhao,~D.; Asmara,~T.~C.; Yu,~X.;
  Breese,~M.~B.; Castro~Neto,~A.~H.; Lam,~Y.~M.; Wee,~A.~T.; Chia,~E.~E.;
  Rusydi,~A. \emph{Physical Review X} \textbf{2018}, \emph{8}\relax
\mciteBstWouldAddEndPuncttrue
\mciteSetBstMidEndSepPunct{\mcitedefaultmidpunct}
{\mcitedefaultendpunct}{\mcitedefaultseppunct}\relax
\EndOfBibitem
\bibitem[Conings \latin{et~al.}(2016)Conings, Babayigit, Klug, Bai, Gauquelin,
  Sakai, Wang, Verbeeck, Boyen, and Snaith]{Conings2016}
Conings,~B.; Babayigit,~A.; Klug,~M.~T.; Bai,~S.; Gauquelin,~N.; Sakai,~N.;
  Wang,~J. T.-W.; Verbeeck,~J.; Boyen,~H.-G.; Snaith,~H.~J. \emph{Advanced
  Materials} \textbf{2016}, 1--9\relax
\mciteBstWouldAddEndPuncttrue
\mciteSetBstMidEndSepPunct{\mcitedefaultmidpunct}
{\mcitedefaultendpunct}{\mcitedefaultseppunct}\relax
\EndOfBibitem
\bibitem[Jeon \latin{et~al.}(2014)Jeon, Noh, Kim, Yang, Ryu, and
  Seok]{Jeon2014a}
Jeon,~N.~J.; Noh,~J.~H.; Kim,~Y.~C.; Yang,~W.~S.; Ryu,~S.; Seok,~S.~I.
  \emph{Nature materials} \textbf{2014}, \emph{13}, 1--7\relax
\mciteBstWouldAddEndPuncttrue
\mciteSetBstMidEndSepPunct{\mcitedefaultmidpunct}
{\mcitedefaultendpunct}{\mcitedefaultseppunct}\relax
\EndOfBibitem
\bibitem[Xia \latin{et~al.}(2016)Xia, Wu, Dong, Xi, Wu, Lei, Xi, Yuan, Jiao,
  Xiao, Gong, and Hou]{Xia2016}
Xia,~B.; Wu,~Z.; Dong,~H.; Xi,~J.; Wu,~W.; Lei,~T.; Xi,~K.; Yuan,~F.; Jiao,~B.;
  Xiao,~L.; Gong,~Q.; Hou,~X. \emph{Journal of Materials Chemistry A}
  \textbf{2016}, \emph{4}, 6295--6303\relax
\mciteBstWouldAddEndPuncttrue
\mciteSetBstMidEndSepPunct{\mcitedefaultmidpunct}
{\mcitedefaultendpunct}{\mcitedefaultseppunct}\relax
\EndOfBibitem
\bibitem[Gao \latin{et~al.}(2018)Gao, Yang, Wang, Sui, Sun, Wei, Cao, and
  Liu]{Gao2018}
Gao,~Y.; Yang,~L.; Wang,~F.; Sui,~Y.; Sun,~Y.; Wei,~M.; Cao,~J.; Liu,~H.
  \emph{Superlattices and Microstructures} \textbf{2018}, \emph{113},
  761--768\relax
\mciteBstWouldAddEndPuncttrue
\mciteSetBstMidEndSepPunct{\mcitedefaultmidpunct}
{\mcitedefaultendpunct}{\mcitedefaultseppunct}\relax
\EndOfBibitem
\bibitem[Jellison and Modine(1996)Jellison, and Modine]{Jellison1996}
Jellison,~G.~E.; Modine,~F.~A. \emph{Applied Physics Letters} \textbf{1996},
  \emph{69}, 371--373\relax
\mciteBstWouldAddEndPuncttrue
\mciteSetBstMidEndSepPunct{\mcitedefaultmidpunct}
{\mcitedefaultendpunct}{\mcitedefaultseppunct}\relax
\EndOfBibitem
\bibitem[Fujiwara \latin{et~al.}(2018)Fujiwara, Kato, Tamakoshi, Miyadera, and
  Chikamatsu]{Fujiwara2018}
Fujiwara,~H.; Kato,~M.; Tamakoshi,~M.; Miyadera,~T.; Chikamatsu,~M.
  \emph{Physica Status Solidi (A) Applications and Materials Science}
  \textbf{2018}, \emph{215}\relax
\mciteBstWouldAddEndPuncttrue
\mciteSetBstMidEndSepPunct{\mcitedefaultmidpunct}
{\mcitedefaultendpunct}{\mcitedefaultseppunct}\relax
\EndOfBibitem
\bibitem[Hirasawa \latin{et~al.}(1994)Hirasawa, Ishihara, and
  Goto]{Hirasawa1994}
Hirasawa,~M.; Ishihara,~T.; Goto,~T. \emph{Journal of the Physical Society of
  Japan} \textbf{1994}, \emph{63}, 3870--3879\relax
\mciteBstWouldAddEndPuncttrue
\mciteSetBstMidEndSepPunct{\mcitedefaultmidpunct}
{\mcitedefaultendpunct}{\mcitedefaultseppunct}\relax
\EndOfBibitem
\end{mcitethebibliography}

\end{document}